\documentclass[aps,preprintnumbers,superscriptaddress,showpacs]{revtex4}
\usepackage{epsfig}
\usepackage{psfrag}
\usepackage{amsfonts}
\usepackage{graphicx}
\usepackage{dcolumn}
\usepackage{bm}

\begin{document}

\title{Entropic destruction of a rotating heavy quarkonium}

\author{Zi-qiang Zhang}
\email{zhangzq@cug.edu.cn} \affiliation{School of mathematics and
physics, China University of Geosciences(Wuhan), Wuhan 430074,
China}

\author{Chong Ma}
\email{machong@cug.edu.cn } \affiliation{School of mathematics and
physics, China University of Geosciences(Wuhan), Wuhan 430074,
China}

\author{De-fu Hou}
\email{houdf@mail.ccnu.edu.cn} \affiliation{Key Laboratory of
Quark and Lepton Physics (MOE), Central China Normal University,
Wuhan 430079,China}

\author{Gang Chen}
\email{chengang1@cug.edu.cn} \affiliation{School of mathematics
and physics, China University of Geosciences(Wuhan), Wuhan 430074,
China}

\begin{abstract}
Using the AdS/CFT duality, we study the destruction of a rotating
heavy quarkonium due to the entropice force in $\mathcal{N}=4$ SYM
theory and a confining YM theory. It is shown that in both
theories increasing the angular velocity leads to decreasing the
entropic force. This result implies that the rotating quarkonium
dissociates harder than the static case.
\end{abstract}

\maketitle
\section{Introduction}
The experiments of ultrarelativistic nucleus-nucleus collisions at
RHIC and LHC have produced a new state of matter so-called quark
gluon plasma(QGP) \cite{JA,KA,EV}. One important signal of the
formation of QGP is quarkonium suppression \cite{TMA}. However,
the recent experimental research of quarkonium production in
nuclear collisions has shown a puzzle: the charmonium suppression
at RHIC (lower energy density) is stronger than that at LHC
(larger energy density) \cite{AAD,BBA}. This is obviously in
contradiction with the Debye screening assumption \cite{TMA} and
the thermal activation scenario \cite{DKH}.

To explain this contradiction, some authors suggest that the
recombination of the produced charm quarks into charmonia may be a
solution \cite{PBR,RLT}. But recently it was argued \cite{DEK}
that this puzzle is related to the nature of deconfinement, based
on the Lattice results \cite{DKA1,DKA2,PPE} which indicate that a
large amount of entropy associated with the heavy quark-antiquark
pair placed in the QGP. It was originally argued in \cite{DEK}
that the entropic force is responsible for dissociating the
quarkonium and this force can be related to the entropy S, that is
\begin{equation}
F=T\frac{\partial S}{\partial L},\label{f}
\end{equation}
where T is the temperature of the plasma.

AdS/CFT \cite{Maldacena:1997re,Gubser:1998bc,MadalcenaReview},
which relates a d-dimensional quantum field theory with its dual
gravitational theory, living in (d+1) dimensions, has yielded many
important insights into the dynamics of strongly-coupled gauge
theories. In this approach, K. Hashimoto et al have first analyzed
the entropic force associated with the heavy quark pair
\cite{KHA}, based on the calculations of the quark-antiquark
potential from AdS/CFT \cite{JMM,ABR,SJR}. It is found \cite{KHA}
that the peak of the entropy near the transition point is related
to the nature of deconfinement. Sooner after \cite{KHA} studies of
the entropic destruction of a moving heavy quarkonium has been
discussed in \cite{KBF}, the authors showed that by increasing the
velocity the moving quarkonium dissociates easier than the static
ones.

In this paper, we extend the holographic studies of \cite{KHA} by
setting the quarkonium to have a angular velocity. We would like
to see how the angular velocity affects the entropic force or the
quarkonium dissociation. It is the motivation of the present work.

The paper is organized as follows. In the next section, we
investigate the entropic force of a rotating quarkonium in
$\mathcal{N}=4$ SYM theory. In section 3, the entropic force of a
rotating quarkonium is studied in a confining YM theory as well.
The last part is devoted to conclusion and discussion.


\section{entropic force of a rotating quarkonium in $\mathcal{N}=4$ SYM theory}

We now analyze the behavior the entropic force associated with a
rotating heavy quark pair in $\mathcal{N}=4$ SYM theory. The
metric is given by
\begin{equation}
ds^2=-\frac{r^2}{R^2}f(r)dt^2+{\frac{r^2}{R^2}}d\vec{x}^2+\frac{R^2}{r^2}\frac{1}{f(r)}dr^2+R^2d\Omega_5^2
\label{metric}, \label{metric}
\end{equation}
where $f(r)=1-\frac{r_h^4}{r^4}$, r denotes the radial coordinate
describing the 5th dimension. R is the AdS radius. The event
horizon is located at $r=r_h$ with $r_h=\pi R^2 T$, where $T$ is
the temperature of the black hole. $d\Omega_5$ is the element of
the solid angle of $S^5$. To consider the rotation, one can
introduce a angular momentum in some $\phi$-direction
\cite{RGC,ABR1,JAC,HXU}. For simplicity, we here consider one
rotational motion direction, i.e. $d\Omega_5=d\phi$.

To proceed, we follow the calculations of \cite{KHA} with the
metric of (\ref{metric}). The Nambu-Goto action is
\begin{equation}
S=T_F\int d\tau d\sigma\mathcal L=T_F\int d\tau d\sigma\sqrt{g},
\end{equation}
where $T_F=\frac{1}{2\pi\alpha^\prime}$ is the fundamental string
tension. $\frac{R^2}{\alpha^\prime}=\sqrt{\lambda}$ with $\lambda$
the 't Hooft coupling. $g$ stands for the determinant of the
induced metric
\begin{equation}
g_{\alpha\beta}=g_{\mu\nu}\frac{\partial
X^\mu}{\partial\sigma^\alpha} \frac{\partial
X^\nu}{\partial\sigma^\beta},
\end{equation}
where $X^\mu$ and $g_{\mu\nu}$ are the target space coordinates
and the metric respectively.

For our assumption, we choose the static gauge,
\begin{equation}
x^0=\tau, \qquad x^1=\sigma,\label{co1}
\end{equation}
and assume that the coordinate $r$ depends on $\sigma$ and the
angular direction $\phi$ depends on $\tau$
\begin{equation}
r=r(\sigma), \qquad \phi=\phi(\tau),\label{co2}
\end{equation}
then the induced metric is found to be
\begin{equation}
g_{00}=\frac{r^2}{R^2}(1-\frac{r_h^4}{r^4})+R^2(\phi^\prime)^2,\qquad
g_{01}=g_{10}=0, \qquad
g_{11}=\frac{r^2}{R^2}+\frac{R^2}{r^2}(1-\frac{r_h^4}{r^4})^{-1}(\dot{r})^2,
\end{equation}
with $\phi^\prime=\frac{\partial\Phi}{\partial\tau}$ and
$\dot{r}=\frac{\partial r}{\partial\sigma}$.

The Lagrangian density $\mathcal{L}$ becomes
\begin{equation}
\mathcal{L}=\sqrt{(\dot{r})^2+\frac{r^4}{R^4}(1-\frac{r_h^4}{r^4})+\frac{R^4}{r^2}(1-\frac{r_h^4}{r^4})^{-1}(\dot{r})^2(\phi^\prime)^2+r^2(\phi^\prime)^2}\label{L}.
\end{equation}

Notice that $\mathcal{L}$ dose not depend on $\sigma$ explicitly,
so the Hamiltonian density is constant, that is
\begin{equation}
H=\mathcal L-\frac{\partial\mathcal
L}{\partial\dot{r}}\dot{r}=constant\label{H}.
\end{equation}

This constant can be found at the special point $r(0)=r_c$ with
$\dot{r}=0$, as
\begin{equation}
H=\sqrt{\frac{r_c^4}{R^4}(1-\frac{r_h^4}{r_c^4})+r_c^2(\phi^\prime)^2}.\label{H1}
\end{equation}

Following (\ref{L}), (\ref{H}) and (\ref{H1}), one has a
differential equation
\begin{equation}
\dot{r}=\frac{dr}{d\sigma}=\sqrt{\frac{a^2(r)-a(r)a(r_c)}{a(r_c)b(r)}}\label{dotr},
\end{equation}
where
\begin{equation}
a(r)=(\frac{r}{R})^4f(r)+r^2(\phi^\prime)^2, \qquad
a(r_c)=(\frac{r_c}{R})^4f(r_c)+r_c^2(\phi^\prime)^2, \qquad
b(r)=1+\frac{R^4}{r^2f(r)}(\phi^\prime)^2,
\end{equation}
with
\begin{equation}
f(r_c)=1-(\frac{r_h}{r_c})^4.
\end{equation}
By integrating (\ref{dotr}) the inter-quark distance $L$ can be
calculated as
\begin{equation}
L=2\int_{r_c}^{r_0}dr\sqrt{\frac{a(r_c)b(r)}{a^2(r)-a(r)a(r_c)}}\label{x}.
\end{equation}
where $r_0=\infty$ is the boundary.

On the other hand, the on-shell action of the fundamental string
in the dual theory is related to the free energy of the quark
anti-quark pair. For small inter-quark distance $L$, the
fundamental string is connected and its on-shell action can be
expressed as
\begin{equation}
F^{(1)}=2T_F\int_{r_h}^{r_0} dr
\sqrt{\frac{a(r)b(r)}{a(r)-a(r_c)}}.
\end{equation}

If the distance $L$ is large enough, the fundamental string will
break in two pieces implying the quarks are screened. For this
case, the free energy is $F^{(2)}$. However, the choice of
$F^{(2)}$ is not unique \cite{MCH}. We here choose a configuration
of two disconnected trailing drag strings \cite{CPH}, that is
\begin{equation}
F^{(2)}=2T_F\int_{r_h}^{r_0} dr.
\end{equation}

To proceed further, we have to resort to numerical methods. In Fig
1, we plot the inter-quark distance $L$ as a function of $r_c/r_0$
at a fixed temperature $r_t/r_0=0.5$ for three different angular
velocity. In the plots from top to bottom
$\phi^\prime=0.8,0.5,0.2$ respectively. One can see clearly that
as $\phi^\prime$ increases the inter-quark distance increases. Or
in other words, the faster the angular velocity, the father the
distance of the heavy quark pair. This can be also understood by
considering the centrifugal force which may have the effect of
increasing $L$.
\begin{figure}
\centering
\includegraphics[width=8cm]{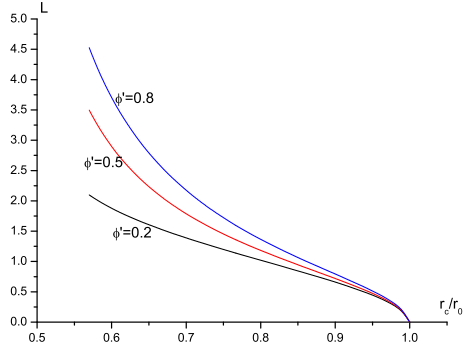}
\caption{The inter-distance $L$ versus $r_c/r_0$ at a fixed
temperature $r_t/r_0=0.5$. Here we take $R=1$.}
\end{figure}

\begin{figure}
\centering
\includegraphics[width=8cm]{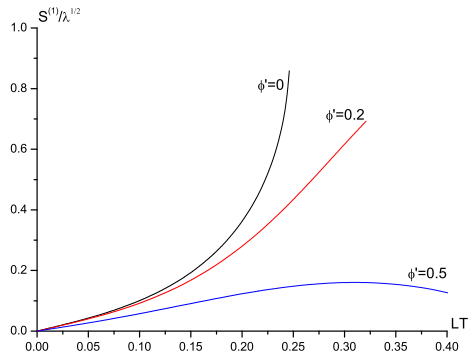}
\caption{The entropy $S^{(1)}/\sqrt{\lambda}$ against $LT$ in
$\mathcal{N}=4$ SYM theory. Here we take $R=1$.}
\end{figure}

\begin{table}
\centering
\begin{tabular}{|c|c|c|c|c|c|c|}\hline
$\phi^\prime$&0&0.01&0.03&0.05&0.08 \\ \hline
c&0.240&0.246&0.249&0.253&0.270\\\hline \multicolumn{3}{c}{}
\end{tabular}
\label{exampletable} \caption{The values of $c$ for some angular
velocity in $\mathcal{N}=4$ SYM theory.}
\end{table}

In addition, the numerical results show that there exist a const
$c$ which is dependent of $\phi^\prime$. If $L>\frac{c}{T}$ the
quarks are completely screened. Here we present the values of $c$
for some different $\phi^\prime$ in table 1. We can see that
increasing $\phi^\prime$ leads to increasing $c$. However, we do
not find the values of $c$ for large $\phi^\prime$. This is
curious.

Next, we calculate the entropy as $S=-\frac{\partial F}{\partial
T}$. For the screened case $L>\frac{c}{T}$, one finds
\begin{equation}
S^{(2)}=\sqrt{\lambda}\theta(L-\frac{c}{T}),
\end{equation}
which implies the entropy at large distance is constant and
independent of the temperature.

For $L<\frac{c}{T}$, to evaluate the entropic force, we study the
growth of the entropy $S^{(1)}$ against the the inter-quark
distance. In Fig.2, we set $R=1$ and plot $S^{(1)}/\sqrt{\lambda}$
as a function of $LT$ with three different $\phi^\prime$. We can
see that  at small distances by increasing the angular velocity
the entropy decreases. Interestingly, if the angular velocity is
large enough, the entropy even decreases as $LT$ increases at
large $LT$.

As stated above, the entropic force, related to the growth of the
entropy with the distance, is responsible for the destruction of
the quarkonium. From the figures, one finds that increasing the
angular velocity leads to decreasing the entropic force. Also, if
the angular velocity is large enough, the entropic force even
becomes "negative" at large $LT$. As a result, we conclude that in
the $\mathcal{N}=4$ SYM theory, the entropic force destructs the
rotating quarkonium harder than the static case.

\section{entropic force of a rotating quarkonium in a confining YM theory}

Next, we investigate the entropic force in a confining YM theory.
To analyze the entropic force around the deconfinement transition,
one should opt for a theory which is confined at low energy and
deconfined at high temperature. The confining SU(N) gauge theory
based on ND4 brans on a circle \cite{EW} satisfies these
conditions. In a deconfined phase, the metric of this theory is
\cite{KHA}

\begin{eqnarray}
ds^2
=(\frac{r}{R})^{3/2}[-f(r)dt^2+(d\vec{x})^2+(dx_4)^2]+(\frac{R}{r})^{3/2}[\frac{1}{f(r)}dr^2+r^2d\Omega_4^2],\label{metric1}
\end{eqnarray}
with
\begin{equation}
f(r)=1-(\frac{r_h}{r})^3,
\end{equation}
where the horizon is fixed at $r=r_h$. The temperature of this
geometry is given by
\begin{equation}
T=\frac{3}{4\pi}\frac{\sqrt{r_t}}{R^{3/2}}.
\end{equation}

Parallel to the $\mathcal{N}=4$ SYM case in the previous section,
we also choose one rotational motion direction as
$d\Omega_4=d\phi$.

The Lagrangian density $\mathcal{L}$ reads
\begin{equation}
\mathcal{L}=\sqrt{(\frac{r}{R})^3f(r)+r^2(\phi^\prime)^2+(\dot{r})^2+\frac{R^3}{rf(r)}(\phi^\prime)^2(\dot{r})^2}\label{L1}.
\end{equation}
with $\phi^\prime=\frac{\partial\Phi}{\partial\tau}$ and
$\dot{r}=\frac{\partial r}{\partial\sigma}$, where we have used
the condition (\ref{co1}) and (\ref{co2}).

We again have a conserved quantity as
\begin{equation}
H=\mathcal L-\frac{\partial\mathcal
L}{\partial\dot{r}}\dot{r}=constant\label{H2}.
\end{equation}

By using the boundary condition at $\sigma=0$,
\begin{equation}
\frac{dr}{d\sigma}=0, \qquad r=r_c,
\end{equation}
we have a differential equation
\begin{equation}
\dot{r}=\frac{dr}{d\sigma}=\sqrt{\frac{a^2(r)-a(r)a(r_c)}{a(r_c)b(r)}}\label{dotr1}.
\end{equation}
where
\begin{equation}
a(r)=(\frac{r}{R})^3f(r)+r^2(\phi^\prime)^2, \qquad
a(r_c)=(\frac{r_c}{R})^3f(r_c)+r_c^2(\phi^\prime)^2, \qquad
b(r)=1+\frac{R^3}{rf(r)}(\phi^\prime)^2,
\end{equation}
with
\begin{equation}
f(r_c)=1-(\frac{r_h}{r_c})^3.
\end{equation}

By solving (\ref{dotr1}), one finds the inter-distance of the
$Q\bar{Q}$ as
\begin{equation}
L=2\int_{r_c}^{r_0}dr\sqrt{\frac{a(r_c)b(r)}{a^2(r)-a(r)a(r_c)}},
\end{equation}
with $r_0=\infty$ the boundary.

Now we proceed to evaluate $F^{(1)}$ and $F^{(2)}$ with numerical
methods, we find that if $L>\frac{c}{T}$ the quarks are screened.
The values of $c$ with respect to $\phi^\prime$ are shown in Table
2.
\begin{table}
\centering
\begin{tabular}{|c|c|c|c|c|c|c|}\hline
$\phi^\prime$&0&0.05&0.1&0.15&0.2 \\ \hline
c&0.252&0.253&0.256&0.261&0.270\\\hline \multicolumn{3}{c}{}
\end{tabular}
\label{exampletable} \caption{The values of $c$ for some angular
velocity in a confining YM theory.}
\end{table}

In the case of $L>\frac{c}{T}$, we find
\begin{equation}
F^{(2)}=2T_F\int_{r_h}^{r_0} dr, \qquad
S^{(2)}=\sqrt{\lambda}\theta(L-\frac{c}{T}).
\end{equation}

For $L<\frac{c}{T}$, we have
\begin{equation}
F^{(1)}=2T_F\int_{r_h}^{r_0} dr
\sqrt{\frac{a(r)b(r)}{a(r)-a(r_c)}}.
\end{equation}

After calculating the entropy $S^{(1)}$ as
$S^{(1)}=-\frac{\partial F^{(1)}}{\partial T}$, we plot
$S^{(1)}/\sqrt{\lambda}$ against $LT$ with three different
$\phi^\prime$ in Fig.3, one can see that the behavior of the
entropy against $LT$ in a confining YM theory is very similar to
the case of $\mathcal{N}=4$ SYM theory, the only difference is the
slope of the curves. In this case, we also find that increasing
the angular velocity leads to decreasing the entropic force at
small distances. Thus, one concludes that in a confining YM
theory, the entropic force destructs the rotating quarkonium
harder than the static case as well.

\begin{figure}
\centering
\includegraphics[width=8cm]{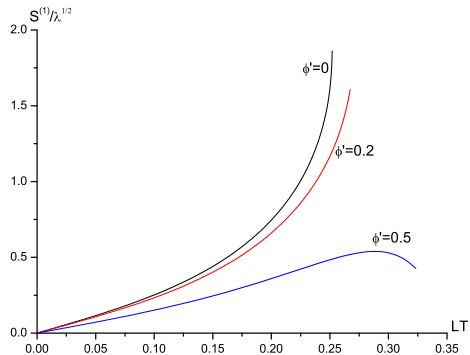}
\caption{The entropy $S^{(1)}/\sqrt{\lambda}$ against $LT$ in a
confining YM theory. Here we take $R=1$.}
\end{figure}

\section{conclusion and discussion}
In this paper, we have investigated the destruction of a rotating
heavy quarkonium due to the entropic force from the AdS/CFT. The
effect of a nonzero angular velocity on the entropic force in
$\mathcal{N}=4$ SYM theory and a confining YM theory has been
studied. It is shown that in both theories, the presence of the
angular velocity tends to decrease the entropic force thus making
the rotating quarkonium dissociates harder than the static case.
To our knowledge, this result is new and different from the
previous studies, see for example in \cite{MAL}.

Interestingly, it was argue \cite{HXU} that increasing the
inter-distance of $Q\bar{Q}$ can be regarded as decreasing the
horizon $r_h$ or decreasing the gravity effect. Since $r_h$ is a
increasing function of $T$, increasing of the angular velocity
leads to decreasing of the system temperature. As we know, lower
system temperature makes the $Q\bar{Q}$ harder to dissociate.
Thus, this agreement also supports that the rotating quarkonium
dissociates harder than the static ones.

Finally, the effect of the rotating heavy quarkonium may be an
explanation to the puzzle on the suppression of the charmonium at
RHIC and LHC: the higher the energy density, the stronger rotating
of the quarkonium, one step further, the smaller the entropic
force, the harder the heavy quarkonium dissociates.

\section{Acknowledgments}

This research is partly supported by the Ministry of Science and
Technology of China (MSTC) under the ¡°973¡± Project no.
2015CB856904(4). Zi-qiang Zhang is supported by the NSFC under
Grant no. 11547204. Gang Chen is supported by the NSFC under Grant
no. 11475149. De-fu Hou is partly supported by NSFC under Grant
nos. 11375070 and 11221504.


\end{document}